\documentclass[prd,twocolumn,amsmath,amssymb,floatfix,superscriptaddress,nofootinbib]{revtex4-1}

\usepackage{graphicx}
\usepackage{scrextend}
\usepackage{amssymb}
\usepackage{amsmath}
\usepackage{bm}
\usepackage{color}
\usepackage{multirow}
\usepackage{cprotect}

\DeclareMathAlphabet\mathbfcal{OMS}{cmsy}{b}{n}
\definecolor{darkgreen}{RGB}{50,150,0}
\definecolor{purple}{cmyk}{0.5,1.0,0,0}

\newcommand{\change}[1]{\textcolor{black}{#1}}

\def\edth{\;\raise1.0pt\hbox{$'$}\hskip-6pt\partial}
\def\baredth{\;\overline{\raise1.0pt\hbox{$'$}\hskip-6pt
\partial}}

\def\be{\begin{equation}}
\def\ee{\end{equation}}
\def\ben{\begin{equation} \nonumber}
\def\een{\end{equation}}
\def\ban{\begin{eqnarray*}}
\def\ean{\end{eqnarray*}}
\def\ba{\begin{eqnarray}}
\def\ea{\end{eqnarray}}
\def\({\left(}
\def\){\right)}

\definecolor{ultramarine}{rgb}{0.07, 0.04, 0.56}
\definecolor{cadmiumgreen}{rgb}{0.0, 0.42, 0.24}
\definecolor{indigo(dye)}{rgb}{0.0, 0.25, 0.42}
\usepackage[linktocpage=true]{hyperref}
\hypersetup{
colorlinks=true,
citecolor=ultramarine,
linkcolor=cadmiumgreen,
urlcolor=indigo(dye),
pdfauthor={},
pdftitle={},
pdfsubject={}
}

\begin{document}

\title{Cross-correlating 2MRS galaxies with UHECR flux from Pierre Auger Observatory }

\author{Pavel Motloch}
\affiliation{Canadian Institute for Theoretical Astrophysics, University of Toronto, M5S 3H8, ON, Canada}

\begin{abstract}
\noindent
We apply a recently proposed cross-correlation power spectrum technique to study
relationship between the ultra-high energy cosmic ray \change{(UHERC)} flux from the Pierre Auger
Observatory and galaxies from the 2MASS Redshift Survey. Using a simple linear bias model
relative to the galaxy auto power spectrum, we are able to constrain  the value of bias to
be less than \change{1\% for UHECR with energies 4 EeV - 8 EeV, less than 2.3\% for UHECR
with energies above 8 EeV and less than 21\% for UHECR with energies above 52 EeV (all
95\% confidence limit).}
We study energy dependence of the bias, but the small
sample size does not allow us to reach any statistically significant conclusions. For the
cosmic ray events above 52 EeV we discover a curious excess cross-correlation at $\sim
1^\circ$ degree scales. Given similar cross-correlation is not visible at larger angular
scales, statistical fluctuation seems like the most plausible explanation.  \end{abstract}

\maketitle

\section{Introduction}
\label{sec:intro}

The quest to find the astrophysical sources of ultra-high energy cosmic rays (UHECR) and
to understand the physical mechanisms that accelerate particles to energies over 1 EeV
is still underway (see \cite{AlvesBatista:2019tlv} for a recent review). Although recently
the Pierre Auger (PA) collaboration achieved a major breakthrough in discovering a dipolar
modulation of the UHECR flux above 8 EeV \cite{Aab:2017tyv}, this does not provide
sufficient information to resolve these puzzles. \change{Smaller scale UHECR anisotropies
would provide further clues, but they so far escape a conclusive detection, despite
several suggestive hints \cite{Abbasi:2014lda,Aab:2018chp}.}

There exists extensive literature dealing with the issue of UHECR anisotropies (see e.g. \cite{
Kashti:2008bw,
Koers:2008ba,
diMatteo:2018vmr,
diMatteo:2020dlo,
Aab:2018chp,
Abbasi:2018tqo,
Tinyakov:2018hfg,
Oikonomou:2012ef,
PierreAuger:2014yba,
Abbasi:2018qlh,
Abbasi:2014lda,
Denton:2014hfa} and references therein). One strategy is to focus exclusively on the UHECR
data. Using them, it is possible to search for ``hotspots''
\cite{Abbasi:2014lda, PierreAuger:2014yba}, constrain individual coefficients in the
spherical harmonic expansion and the related angular power spectra of map of the UHECR
flux \cite{diMatteo:2018vmr, Denton:2014hfa}
or investigate the correlation function \cite{Tinyakov:2018hfg,
PierreAuger:2014yba}.
Alternatively, one can search for a connection between the observed directions of the UHECR events
and the distribution of objects in the nearby Universe. Beyond
again looking at the correlation function \cite{diMatteo:2020dlo, PierreAuger:2014yba}, it is also
possible \cite{Kashti:2008bw,Koers:2008ba,Oikonomou:2012ef, Tinyakov:2018hfg,
diMatteo:2018vmr, Aab:2018chp, PierreAuger:2014yba,Ahlers:2017wpb} to build an expected UHECR
flux based on a particular physical model for the sources and the
magnetic fields and compare the observed distribution of UHECR with the expectation of the
model, for example using the Kolmogorov-Smirnov statistic or by comparing likelihood
ratios of various models. 

Recently, \cite{Urban:2020szk} proposed using harmonic space cross-correlation power
spectrum (see Eq.~\ref{cell}) of the UHECR flux and a tracer of the large scale structure
as a novel addition to the later family of approaches. Effectively, it compresses the two
dimensional information contained in the two maps into a one dimensional cross power
spectrum $C_\ell^{XY}$ that quantifies coherence of features in the maps as a function of
the angular scale. Such compression would be detrimental for example in case the
UHECR are produced by several nearby sources and information on the map level would be
expected to be essential. On the other hand, in the case of a large number of sources tracing
the large scale structure, information about any particular sky direction is not important
and it is mostly the overall coherence between UHECR and the tracer that carries the
relevant information. In such case reducing the dimensionality of the data can be
advantageous. Even if we find that the UHECR flux is dominated by several nearby sources,
analysis based on the cross-correlation power spectra might be useful to understand the
properties of the residual flux after these sources are modeled and subtracted out.

The usefulness of the cross-power spectrum analysis also depends on the effect the
magnetic fields have on the UHECR propagation.  If the magnetic fields cause mostly a
stochastic deflection of UHECR, this should manifest effectively as smearing of the signal
below a certain smoothing scale.  The cross-power spectrum technique would then still be
useable on angular scales larger than such smoothing scale, with the value of the
smoothing scale providing information about the properties of the magnetic field. On the
other hand, if the effects of the magnetic field turn out to be more complicated, map based
analysis may again turn out to be necessary.

Notice that the cross-correlation technique does not depend on any particular physical
model (of for example the magnetic field) beyond choosing the particular tracer to
cross-correlate with. This means that it is well suited for initial explorations, findings
of which can inform more complex physical models.

While the proposed technique is at heart related to the other commonly used
techniques (for example the cross-power spectrum is a Fourier transform of the real space
correlation function), it offers certain advantages. In particular, it offers a clean
separation of angular scales, which may offer additional insights. In commonly used
methods similar information could be obtainable by comparing results obtained with
different smoothing scales, but this can quickly become rather involved. 

In this work we perform the first application of the cross-correlation power spectrum
technique to a UHECR data set, namely events detected by the PA Observatory (PAO). We
cross-correlate the corresponding UHECR flux with galaxies from the 2MASS Redshift Survey
(2MRS); by doing this we effectively probe how much do the UHECR events trace all the
visible matter in the nearby Universe. As a possible alternative one can perform a targeted
analysis by picking a particular set of objects deemed likely to contribute to the UHECR
flux (such as starburst galaxies) and cross correlate the UHECR flux with them, but we do
not perform such a targeted analysis in this work.

The ultimate question we will aim to answer is whether there is any excess correlation
between the 2MRS and PAO datasets. While the cross-correlation does not need to assume any
particular model (see the points in Fig.~\ref{fig:cross}), we fit a simple linear bias
model relative to the 2MRS galaxy auto power spectrum to gauge to what extent are the data consistent
with no correlation.
To the extent allowed by the publicly available data we will also investigate dependence
of the bias on the energies of the UHECR events.

This work is organized as follows: We start by summarizing the data sets we use in
\S~\ref{sec:data}. In \S~\ref{sec:analysis} we describe the steps used in our analysis,
before presenting our results in \S~\ref{sec:results}. We list tests we performed to
check robustness of our results in \S~\ref{sec:tests} and conclude with a discussion in
\S~\ref{sec:discuss}. In the Appendix~\ref{sec:app} we describe how we calculate the
exposure of the Pierre Auger Observatory.

\section{Data}
\label{sec:data}

We use the PAO data publicly released with \cite{Aab:2017tyv}, representing over 12 years
of observing. We mostly focus on the 32187 UHECR events with energies above 8 EeV, although we
briefly discuss also the 81701 events with energies between 4 and 8 EeV. 
Beyond this binary division we do not have any further information about the
energies of the events, with the exception of a subset of 231 events with energies above
52 EeV which were released earlier and which correspond to about 10 years of data
collecting \cite{PierreAuger:2014yba}. For a technical description of the
PAO and the detection techniques, we refer the reader to \cite{ThePierreAuger:2015rma}.

As a tracer of the nearby large scale structure we use galaxies from the 2MRS
\cite{Huchra:2011ii}. This spectroscopic survey covers 91\% of sky and contains 43533
galaxies, forming a nearly complete (97.6\%) survey of galaxies brighter than 11.75 mag in
$K_s$ band, with reddening limited to $E(B-V) \le 1\, \mathrm{mag}$.

\section{Analysis techniques}
\label{sec:analysis}

In this section we go over the technical details of our analysis. We first describe how we
convert the catalogs of UHECR events and galaxies into sky maps, how we estimate their cross
power spectra and introduce our linear bias model. Then we go on and detail \change{the}
generation of mock UHECR catalogs, comment on obtaining auto power spectra of the 2MRS
galaxy catalog and finish by explaining a maximum likelihood estimate for the bias
parameter of our model.

\subsection{Maps and masks}
\label{sec:maps}

We work with maps in the Healpix pixelization \cite{Gorski:2004by}. Our default resolution
will be $N_\mathrm{side} = 512$, with each pixel having an area of 47 square
arc minutes.

From the 2MRS data, we construct the galaxy overdensity map
\be
\label{gal_overdensity}
	\delta^g(\vec n) = \frac{n_g(\vec n) - \bar n_g}{\bar n_g} ,
\ee
where $n_g$ denotes the number of galaxies in a given pixel and $\bar n_g$ is the mean value
over all observed pixels. Only galaxies with galactic latitude \change{$b$} and longitude
\change{$l$} satisfying
\be
	|b| \ge
	\begin{cases}
	5^\circ, \ \ \ 30^\circ \le l \le 330^\circ\\
	8^\circ, \ \ \ \text{otherwise}\\
	\end{cases}
\ee
were included into the catalog; this determines the mask of the survey which we will use
in what follows.

From the UHECR catalog we first convert the observed hit counts into a flux map $\Phi$ by
using an estimated exposure function \eqref{exposure_func} from Appendix~\ref{sec:app}.
Then we calculate the overdensity map $\delta^\Phi$ in a manner analogous to
\eqref{gal_overdensity}. The mask is determined by the maximal declination that can be
observed by PAO.

\subsection{Power spectra and linear bias}
\label{sec:power}

Given two fields on a sphere $\delta^X, \delta^Y$, their cross power spectrum is
calculated as 
\be
\label{cell}
	\hat C^{XY}_\ell = \sum_m \frac{\delta^X_{\ell m}\delta^{Y*}_{\ell m}}{2 \ell + 1} ,
\ee
where $\delta^{X,Y}_{\ell m}$ are coefficients of $\delta^{X,Y}$ in the spherical
harmonic expansion and star denotes complex conjugation. For brevity, in what follows we
will not explicitly write the hat on top of power spectra to denote an estimate.

The model we will consider in this work is a linear relationship between the cross-power
spectra of $\delta^\Phi$ and $\delta^g$, the  UHECR flux and galaxy counts overdensities,
and the galaxy overdensity auto power spectrum,
\be
\label{model}
	C^{\Phi g}_{\ell} = b C^{gg}_\ell + N_\ell.
\ee
Here $N_\ell$ represents noise and $b$ is commonly referred to as
``bias''. In this work we will estimate value of $b$ to see whether it is consistent with
zero, which would mean no relationship between the anisotropies in the observed UHECR flux
and the 2MRS galaxies.

Because of the incomplete sky coverage, we do not know $\delta^X$ over the whole
sky. Effectively we can think of this as measuring $\delta^X$ modulated with a
position-dependent weight $w^X$, i.e.
\be
	\tilde \delta^{g}(\vec n) = w^g(\vec n) \delta^{g}(\vec n) .
\ee
In the simplest case, $w^X$ represents a binary mask and is set to zero in parts of the sky
not accessible to the experiment and to one elsewhere. \change{In more complicated cases, $w^X$ can
be used to down-weight the data in parts of the sky where the measurements are relatively
noisier. The unobserved pixels are the limiting case, as they can be thought of as being
infinitely noisy.}

We can then calculate the cross-power spectra of $\tilde \delta^{g}, \tilde
\delta^{\Phi}$, so-called pseudo-power spectra $\tilde C^{g\Phi}_\ell$.
Assuming no mixing between different angular scales, these are related to the underlying
power spectra through a linear relation (e.g. \cite{Hivon:2001jp})
\be
	\tilde C^{g\Phi}_\ell = \sum_{\ell'} K_{\ell \ell'} C^{g\Phi}_{\ell'} ,
\ee
where the kernel $K_{\ell \ell'}$ can be analytically calculated from the weights $w^X$.
For our weights the kernels will be invertible, which allows us to express
\be
	C^{g\Phi}_\ell = \sum_{\ell'} K_{\ell \ell'}^{-1} \tilde C^{g\Phi}_{\ell'} .
\ee
To calculate the pseudo-power spectra and the kernel $K_{\ell \ell'}$ we use the publicly
available code Polspice \cite{Chon:2003gx}.

While for the galaxy overdensity map we use $w^g$ based on the 2MRS mask, for the PA
events we use the product of the corresponding mask and the exposure function $E(\delta)$
described in Appendix~\ref{sec:app}. This way we put lower weight on events in the
parts of the sky with low exposure, that are intrinsically more noisy. We point out that
\change{given the underlying Poisson statistics,}
the choice of weighting of the PAO events leading to the optimal signal to noise
ratio is unknown to us and does not appear to be immediately obvious, so we use this
as a simple heuristic.

\subsection{Random UHECR event catalog}

Here we describe how we generate mock UHECR catalogs, necessary to estimate properties of
$N_\ell$ and uncertainties of our analysis. As we will see shortly, the
values of $b$ preferred by the data are rather small and we are not able to rule out $b =
0$ with sufficient statistical significance. We will thus simulate UHECR events as if
they were independent of the galaxy positions.

We start by generating events from an isotropic flux, consistent with the exposure
\eqref{exposure_func}. To accommodate for the dipole observed in the UHECR
flux \cite{Aab:2017tyv}, for each event we first calculate
\be
	p_\mathrm{th} = \frac{1 + A_d\(\vec n \cdot \vec n_\mathrm{PAd}\)}{1 + A_d} ,	
\ee
where $\vec n$ is the direction of the generated event, $\vec n_\mathrm{PAd}$ is the
direction of the observed UHECR flux dipole (declination $-24^\circ$ and right ascension
$100^\circ$) and $A_d = 0.065$ is its amplitude. For each such event we then draw a
random number $p$ uniformly distributed in $[0, 1]$ and retain the generated event only if
$p < p_\mathrm{th}$; this leads to a dipole flux modulation. We repeat this procedure
until we have the same number of events as is in the UHECR catalog, and we use them to
construct a mock overdensity map $\delta^\Phi$. The whole process is repeated to generate
$N_\mathrm{sims} = 200$ mock UHECR maps.

\subsection{2MRS auto power spectra}

To estimate the value of bias $b$ we also need to know the auto power spectrum of the
2MRS galaxy overdensities. To achieve this, we repeat the calculation from
\cite{Ando:2017wff}.

Because the angular power spectrum estimated from $\delta^g$ by Polspice is a sum of the
signal and shot noise, we must subtract the latter. To estimate it, we randomly split the
2MRS galaxy catalog into two parts, obtain the corresponding overdensities
$\delta^{g,1}, \delta^{g,2}$ and form the half-sum and half-difference maps
\ba
	HS &=& \frac{\delta^{g,1} + \delta^{g,2}}{2}\\
	HD &=& \frac{\delta^{g,1} - \delta^{g,2}}{2} .
\ea
As the latter contains on average only the shot noise and no signal, we can estimate the
2MRS galaxy power spectrum as
\be
	C^{gg}_\ell = C^{HS}_\ell - C^{HD}_\ell .
\ee
The power spectrum we get this way is in good agreement with values reported in
\cite{Ando:2017wff}. 

\subsection{Fitting $b$}

The power spectra $C^{g\Phi}$ obtained from the mock UHECR maps reveal that 
the expectation value of $N_\ell$ is, while small, nonzero \change{(mostly because the generated
UHECR events have the dipolar modulation)}. We calculate its mean value
$\langle N_\ell\rangle$ from simulations, and define
\ba
	\Delta C^{\Phi g}_{\ell} &=& C^{\Phi g}_{\ell} - \langle N_\ell \rangle\\
	\Delta N_\ell &=& N_\ell - \langle N_\ell \rangle .
\ea
We will then try to explain the \emph{excess} correlation between UHECR and galaxies in
the model,
\be
\label{model_unbiased}
	\Delta C^{\Phi g}_{\ell} = b C^{gg}_\ell + \Delta N_\ell,
\ee
now with noise that has a zero expectation value.

To avoid correlations and have a more Gaussian likelihood, we bin the multipoles with
$\Delta \ell = 21$, starting with $\ell_\mathrm{min} = 2$. As the resolving power of PAO
at the investigated energies is $\sim 1^\circ - 2^\circ$, we consider either five
($\ell_\mathrm{max} = 106$) or ten ($\ell_\mathrm{max}
= 211$) bins in our analysis, as these are the approximate $\ell$ ranges that correspond
to the angular scales observed. We chose these $\ell_\mathrm{max}$ before the analysis, to
avoid a posterior bias.

On simulations we checked that the binned $\Delta C^{\Phi g}_\ell$ are well described by a
Gaussian distribution with a diagonal covariance; these conclusions are not expected to be
affected if $b$ turns out to be nonzero but small.

These findings allow us to construct a Gaussian likelihood $\mathcal{L}$ for $b$,
\be
\label{likelihood}
	- 2\log \mathcal{L} = \sum_B \(\Delta C^{\Phi g}_B - b C^{gg}_B\)^2\sigma_B^{-2} ,
\ee
where $B$ sums over bins and $\sigma_B$ is standard deviation of $\Delta C^{\Phi g}_B$
estimated from the simulations.

\change{Taking the derivative of the right hand side of \eqref{likelihood} with respect to
$b$, setting it equal to zero and solving for $b$ leads to the maximum likelihood
estimator}
\be
	b = \frac
	{
	\sum_B \Delta C^{\Phi g}_B C^{gg}_B\sigma_B^{-2}
	}
	{
	\sum_B \(C^{gg}_B\)^2\sigma_B^{-2}
	}.
\ee
Its standard deviation \change{can be read off directly from the $b^2$ term in
\eqref{likelihood} and is equal to}
\be
	\sigma_b^{-2} = \sum_B \(C^{gg}_B\)^2\sigma_B^{-2} .
\ee

\begin{figure}
\center
\includegraphics[width = 0.49 \textwidth]{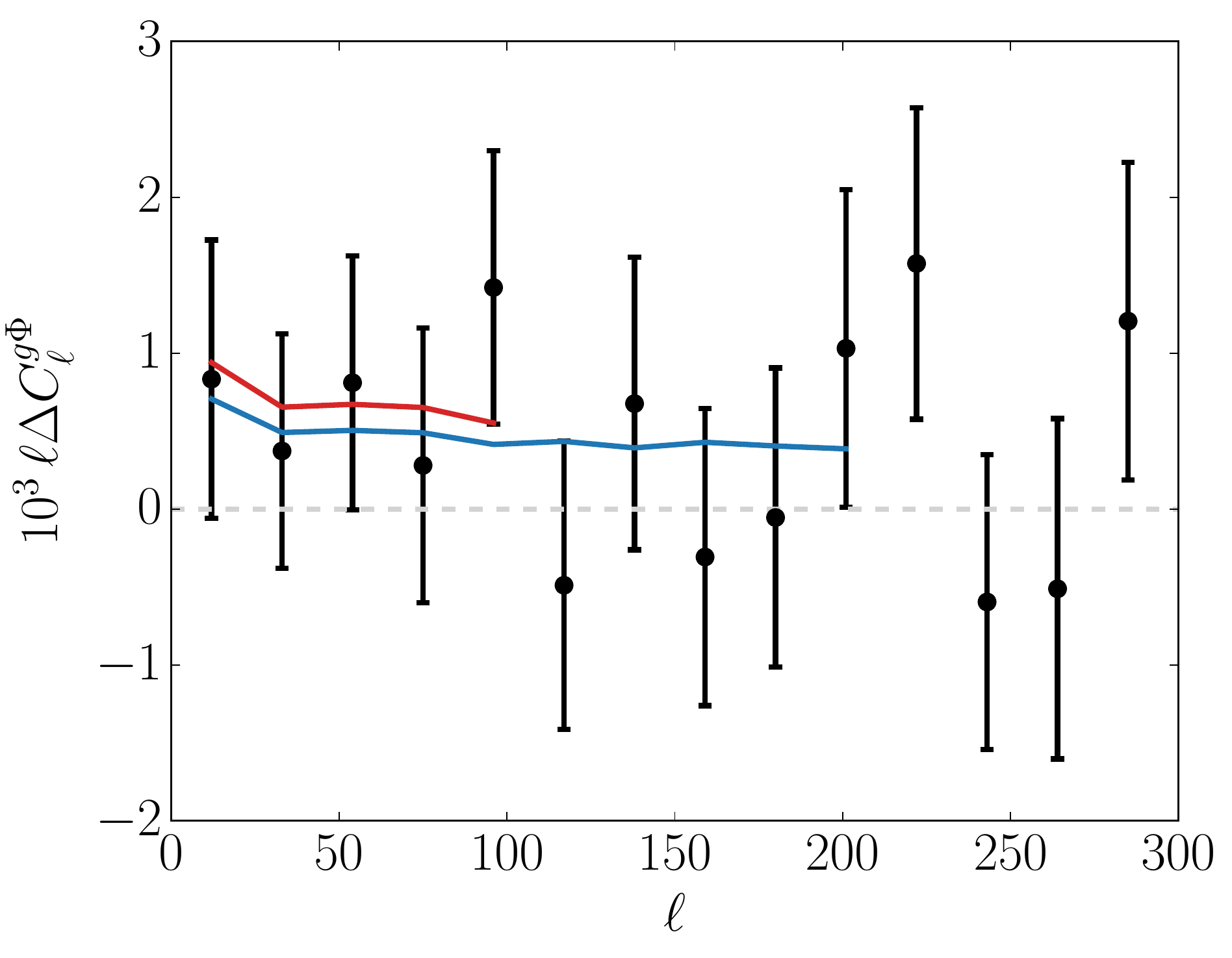}
\caption{Excess cross-correlation (black points) between the overdensities of the UHECR flux ($E >
8\,\mathrm{EeV}$) and galaxy count,
together with standard deviations estimated from simulations. The dashed line corresponds
to the $b=0$ model, the red and blue line show the best fit linear bias model $bC^{gg}$
with $b$ estimated using data to $\ell_\mathrm{max}$ of 106 and 211 respectively.}
\label{fig:cross}
\end{figure}

\begin{figure}
\center
\includegraphics[width = 0.49 \textwidth]{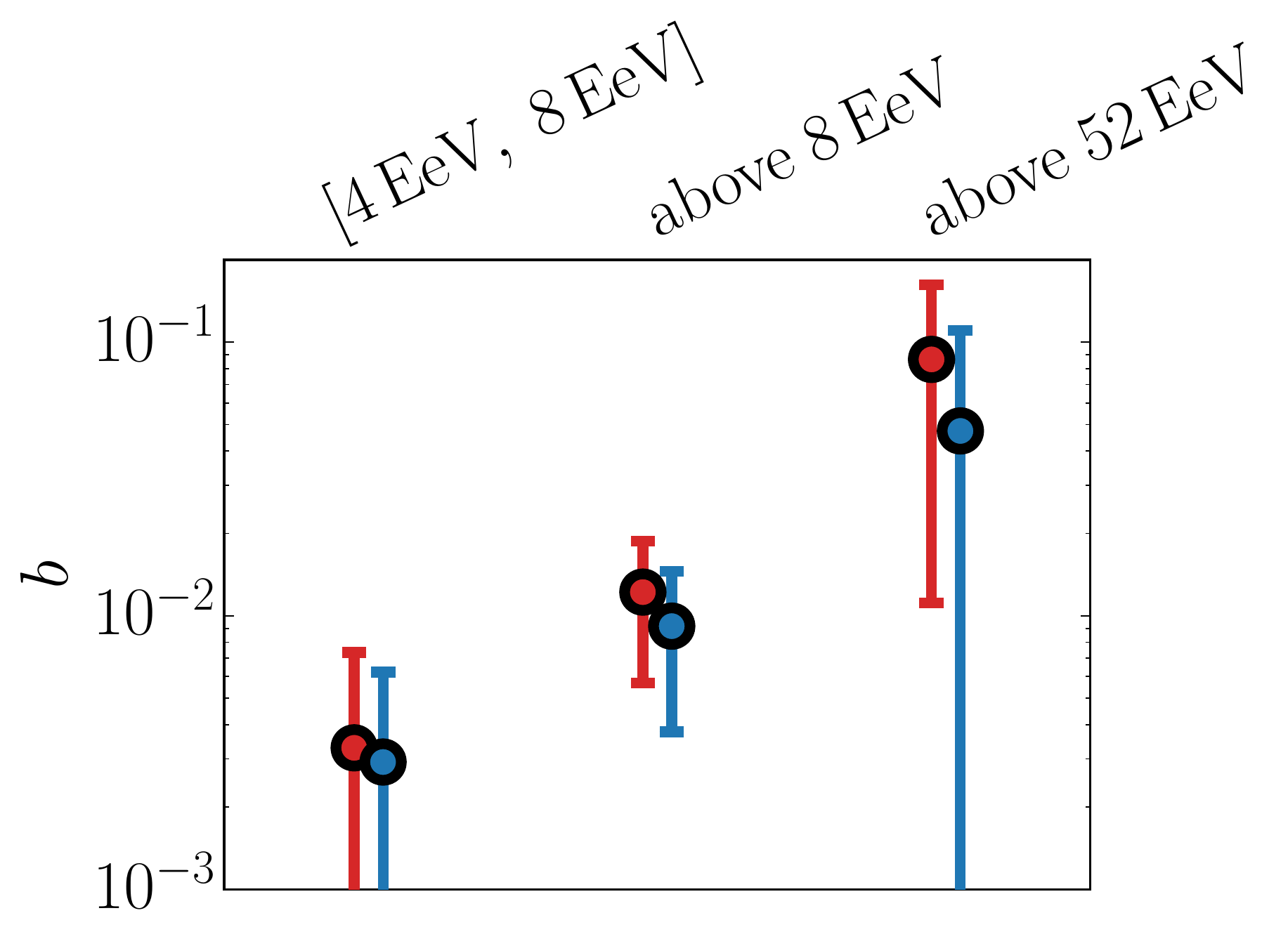}
\caption{The best fit bias $b$ in three UHECR energy bins. Using the events between 4 EeV and 8
EeV (left), our fiducial sample with energies above 8 EeV (center) and the sample of events with energies
above 52 EeV (right, overlaps the fiducial sample).
Estimated using data to $\ell_\mathrm{max}$ of 106 (red) and 211
(blue).
}
\label{fig:energy}
\end{figure}

\begin{table}
\caption{Best fit values of bias} 
\label{tab:b}
\begin{tabular}{ccc}
\hline\hline
UHECR Energy & \multicolumn{2}{c}{$\ell_\mathrm{max}$}
\\
& 
$106$ 
&	
$211$ 
\\
\hline
4 EeV -- 8 EeV &
$\(0.33 \pm 0.41\)\times 10^{-2}$
& 
$\(0.29 \pm 0.33\)\times 10^{-2}$
\\
above 8 EeV &
$\(1.22 \pm 0.65\)\times 10^{-2}$
& 
$\(0.92 \pm 0.54\)\times 10^{-2}$
\\
above 52 EeV &
$\(8.65 \pm 7.54\)\times 10^{-2}$
& 
$\(4.74 \pm 6.29\)\times 10^{-2}$
\\
\hline\hline
\end{tabular}
\end{table}

\section{Results}
\label{sec:results}

In Fig.~\ref{fig:cross} \change{we show the excess cross-correlation between galaxy and
UHECR positions for our fiducial sample of
UHECR events with energies over 8 EeV}. The case of no bias, $b = 0$, is represented by the
dashed line, the red and blue curves represent the best fit $b C^{gg}$ when
$\ell_\mathrm{max}$ is either 106 or 211. The best fit values of $b$ are listed in
Table~\ref{tab:b}, we see a $\sim 1.8\sigma$ detection of positive bias of about a
percent. The best fit bias is slightly smaller when the larger range of multipoles
$\ell_\mathrm{max}$ is used.

We repeat the analysis for the UHECR events with energies between 4 EeV and 8 EeV and for
the subset of events we know have energies above 52 EeV and show the resulting values of
$b$ in Fig.~\ref{fig:energy}. 
The best fit values of $b$ are again listed in Table~\ref{tab:b}.

\begin{figure*}
\center
\includegraphics[width = 0.99 \textwidth]{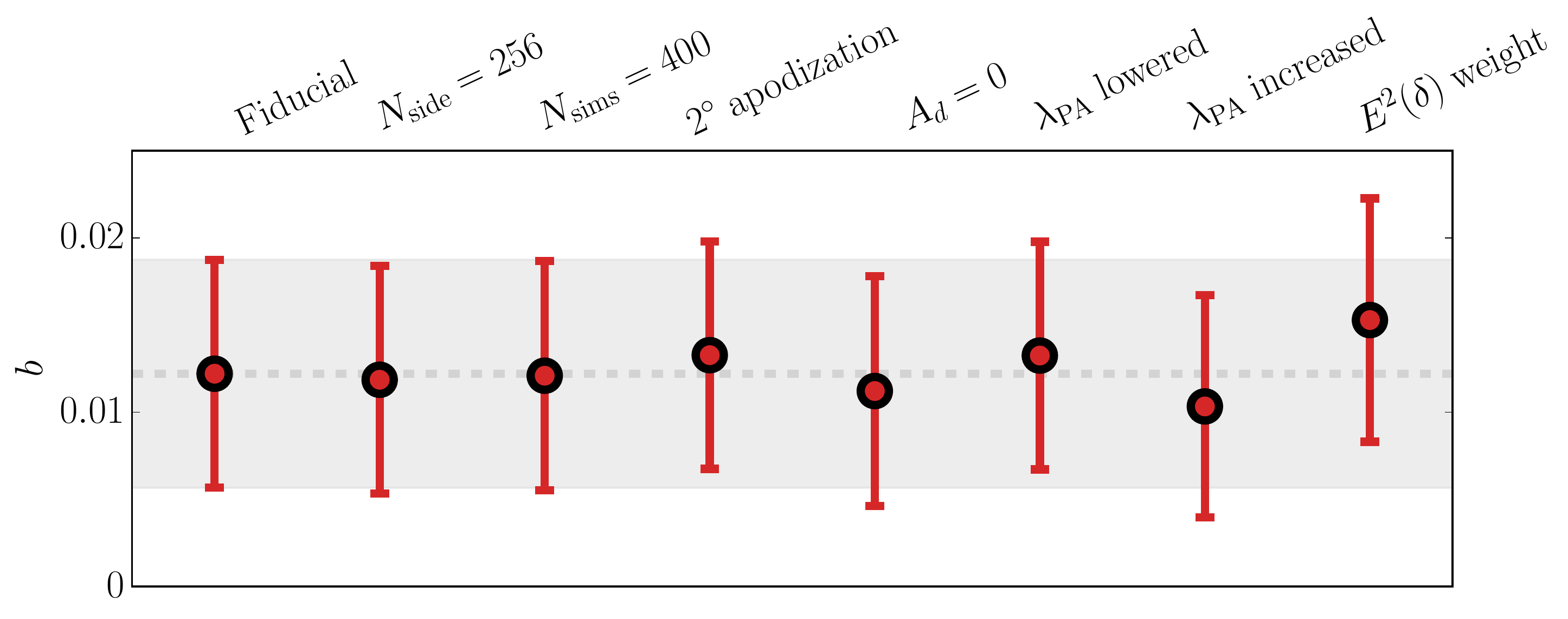}
\caption{Testing impact of various assumptions (see text) on the outcome of our analysis,
with $\ell_\mathrm{max} = 106$. The left-most point is our fiducial analysis.}
\label{fig:tests}
\end{figure*}

\begin{figure}
\center
\includegraphics[width = 0.49 \textwidth]{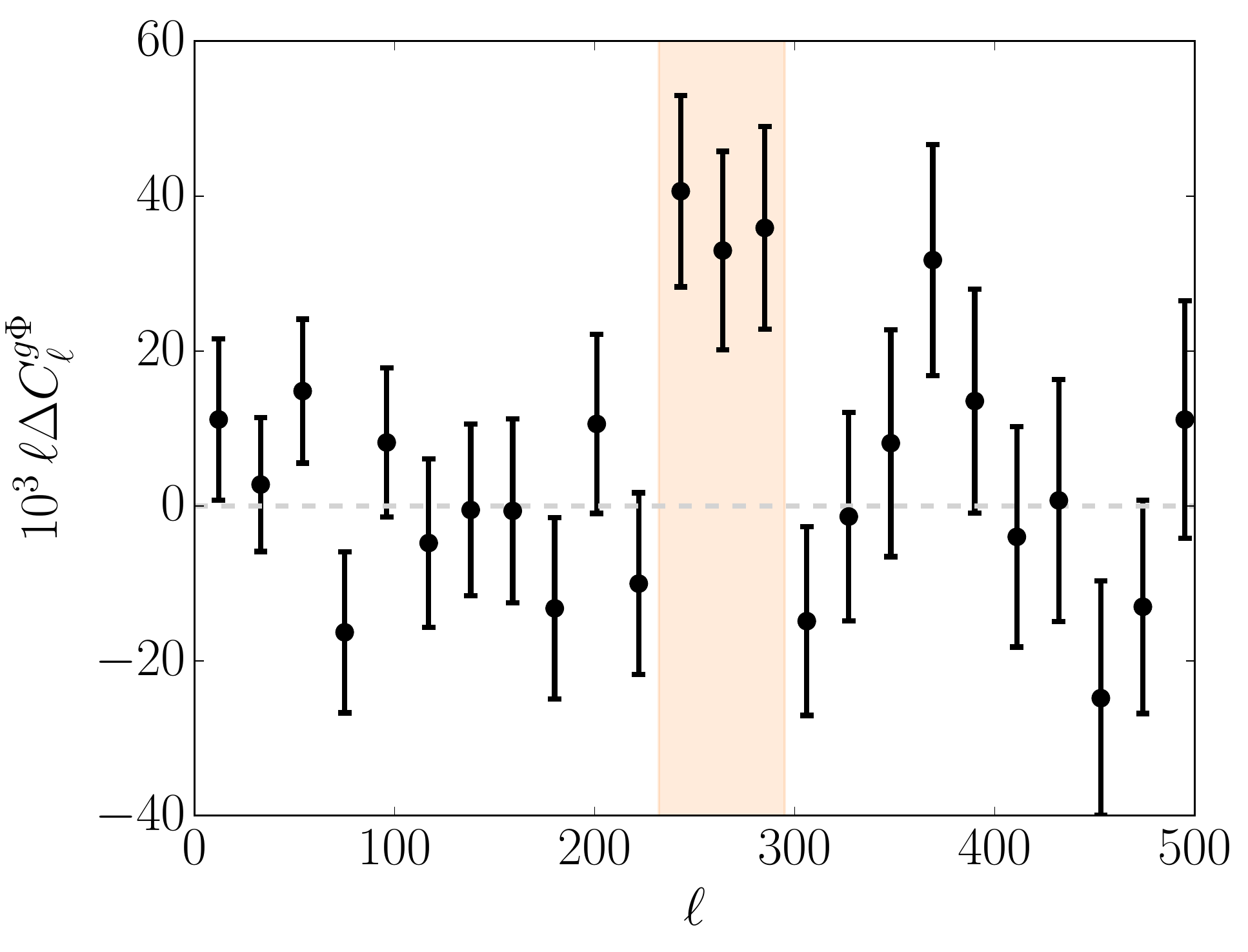}
\caption{Excess cross-correlation between the overdensities of the flux of UHECR events
with energies above 52 EeV and galaxy counts, together with standard deviations estimated
from simulations (black points). 
The orange range shows the region with surprisingly strong cross-correlation.
}
\label{fig:cross_highE}
\end{figure}

For the flux of UHECR events with energies above 52 EeV we noticed a strange excess correlation
in the multipole range 235 - 295, see Fig.~\ref{fig:cross_highE}. Local significance of
the cross correlations observed in these three bins are 3.3$\sigma$, 2.6$\sigma$ and
2.7$\sigma$, which together with their clustering we found interesting enough to report
here.

\section{Tests}
\label{sec:tests}

For the catalog of events with energies above 8 EeV we repeated the analysis several times
with different analysis choices, to test convergence and sensitivity to our assumptions. In
Fig.~\ref{fig:tests} we show the final values of the
bias for the case with $\ell_\mathrm{max} = 106$; $\ell_\mathrm{max} = 211$ leads to
similar conclusions.

We tested:
\begin{itemize}
	\item halving $N_\mathrm{side}$, the resolution of the underlying Healpix grid
	\item doubling $N_\mathrm{sim}$, the number of Monte Carlo simulations
	\item apodizing the masks with a $2^\circ$ cosine taper to ascertain there are no
	problems with ringing in the Fourier domain
	\item neglecting the UHECR dipole when generating mock catalogs by setting $A_d = 0$
	\item changing $E(\delta)$ by artificially increasing / lowering the PAO latitude
	$\lambda_\mathrm{PAO}$ (see Appendix~\ref{sec:app}) while keeping the observable part
	of the sky constant through a matching change in $\theta_\mathrm{max}$
	\item weighting the PAO events by $E^2(\delta)$ as opposed to $E(\delta)$ (see
	\S~\ref{sec:power})
\end{itemize}

We also tested that choosing a finer binning with $\Delta \ell = 5$ or using the full
covariance matrix --- as opposed to just the diagonal elements --- to estimate $b$ does
not significantly change the results.

We performed a similar suite of tests (without $N_\mathrm{side}$ and apodization) for the
UHECR catalog with energies over 52 EeV and found that the $\sim 1^\circ$ anomaly is
similarly robust with respect to our analysis choices.

\section{Discussion}
\label{sec:discuss}

While the currently available data do not allow for any statistically significant
detection, for the UHECR with energies above 8 EeV we see a preference for $b \sim 0.01$
with about a 1.8$\sigma$ random chance probability. If this correlation remains as more data
is collected, this would provide an evidence that at least part of the UHECR events
is related to the 2MRS galaxies. Interpreting our result from the opposite point of view,
we limit the allowed values of $b$ for this sample to below \change{0.023 (95\% confidence
limits).}
\change{
Although we find preference for positive $b$ in all energy bins we study, in the other
energy bins we do not observe nonzero $b$ with such a high significance. This is not
surprising, given that in the lower energy bin PAO is yet to detect the UHECR dipole and that in
the highest energy bin the sample size is limited, which leads to significantly increased
error bars.
}

In general, we find a systematic decrease in the best fit value of $b$ when
using a larger range of multipoles $\ell$ in the estimate. This is not surprising, as the high
$\ell$ data are expected to be increasingly noise dominated due to the PAO resolution. 

A natural explanation for the low value of the observed bias is a strong deflection of
the UHECR by the magnetic fields between the UHECR sources and Earth, erasing the
relationship between the directions to the source and the observed UHECR direction. A straightforward
test of this hypothesis is investigation of the dependence of the bias on the UHECR
energy, with the expectation of bias growing with energy as the high energy UHECR are less deflected.
By looking at events in the 4 EeV -- 8 EeV energy range and events above 52 EeV we indeed
see a hint of the bias rising as the energy of UHECR sample increases (Fig.~\ref{fig:energy}),
but there is insufficient statistic to make any claims. It would be interesting to split
the PA events
above 8 EeV into finer energy bins and repeat the analysis, but to our knowledge the
energies of individual events (except for the small number of events released with
\cite{PierreAuger:2014yba}) are not public.

Alternative explanation of the low values of $b$ would be that the UHECR trace not the
full large scale structure as expressed through the 2MRS galaxies, but only a small subset
of the galaxies (e.g. only the starburst galaxies). If this is the case, then
cross-correlating with the whole 2MRS catalog effectively dilutes the signal, leading to
lower values of $b$. Another avenue for future research is thus looking into correlations
with more specific large scale structure datasets. Alternatively, one can consider linear
combinations of multiple tracers, each with its own individual bias factor $b_i$, or split
the objects to cross-correlate with into redshift bins to investigate redshift dependence.

The curious excess cross correlation between the highest energy UHECR sample and the 2MRS galaxies
\change{(see Fig.~\ref{fig:cross_highE})}
appears at scales that correspond to slightly below
a degree or so, $\ell$ in the range 235 -- 295. 
While at these scales we would expect PAO to quickly start losing sensitivity due to
experimental resolution, in principle PAO might still be capable of picking up signal here.
However, it is hard to imagine a physical mechanism that would
cause correlations at these scales which would not also manifest itself on larger angular
scales. 
\change{Probability to observe three consecutive bins fluctuate up by combined 8.6 standard 
deviations, which is what we observe in the data, is $5 \cdot 10^{-6}$ when considering 15
bins ($\ell_\mathrm{max} = 316$) and $7 \cdot 10^{-6}$ when considering 20 bins
($\ell_\mathrm{max} = 421$).
However, one should be careful not to take these significances at face value as the anomaly
was
discovered only a posteriori. For example, it might have been that an anomaly appeared in
two or four bins, instead of three, or had a more complicated $\ell$ dependence. It is not
clear how exactly to calculate probability that ``an anomaly'' appears in the data
and one should keep this in mind when interpreting the probabilities quoted in this
paragraph.
}
Fortunately, PAO should have detected over 100 new events with energies above 52
EeV that are not included in \cite{PierreAuger:2014yba}. With these new events, it should
be possible to quickly confirm
that this anomaly is just a statistical fluctuation, now without any penalty for an a posteriori
selection. 

Finally, we want to stress how the cross power spectrum technique allows us to simply
probe the angular and energy dependence of the relation between the UHECR flux and the nearby large scale
structure, something that is often clouded by scanning over energy and a smoothing scale
in a typical anisotropy search.

\acknowledgements{
We thank Federico R. Urban for useful discussions and an anonymous referee for useful
suggestions.
}

\appendix

\section{Pierre Auger Observatory Exposure}
\label{sec:app}

To approximate the exposure of the Pierre Auger Observatory, located at latitude
$\lambda_\mathrm{PAO} = -35.21^\circ$, as a function of declination, we will assume PAO is
capable of detecting all UHECR events coming from zenith angles $\theta <
\theta_\mathrm{max} = 80^\circ$, which was the cutoff selected in \cite{Aab:2017tyv,PierreAuger:2014yba}.

Working in a coordinate system with origin placed in the Earth's center, with the $z$ axis
going through the north pole and PAO located on the $x$ axis, the position of PAO is
\be
	\vec v_\mathrm{PAO} = \(\cos \lambda_\mathrm{PAO}, 0, \sin \lambda_\mathrm{PAO}\) .
\ee

We can parameterize a general UHECR event as coming from 
\be
	\vec v_e = \(\cos \delta \cos \phi, \cos \delta \sin \phi, \sin \delta\) ,
\ee
where $\delta$ is the declination and $\phi \in [-\pi, \pi]$. To be detected, this event
must satisfy
\be
	\vec v_e \cdot \vec v_\mathrm{PAO} > \cos \theta_\mathrm{max} ,
\ee
which is equivalent to
\ba
	\cos \phi
		&>& \frac{
		\cos \theta_\mathrm{max} - \sin \delta \sin \lambda_\mathrm{PAO}
		}{\cos \delta \cos \lambda_\mathrm{PAO}}\\
	|\phi| &<& \phi_\mathrm{cr}(\delta) ,
\ea
where for our values of $\lambda_\mathrm{PAO}$
\be
	\phi_\mathrm{cr}(\delta) =
	\begin{cases}
		0,\ \ \delta > \theta_\mathrm{max} + \lambda_\mathrm{PAO}\\
		\pi,\ \ \delta < \theta_\mathrm{max} - \lambda_\mathrm{PAO}-\pi\\
		\arccos\(
		\frac{
		\cos \theta_\mathrm{max} - \sin \delta \sin \lambda_\mathrm{PAO}
		}{\cos \delta \cos \lambda_\mathrm{PAO}}
		\), \ \ \text{otherwise} .
	\end{cases}
\ee

The exposure for given declination is then proportional to
\ba
\label{exposure_func}
	E(\delta) &\propto& 
	\int_{-\phi_\mathrm{cr}(\delta)}^{\phi_\mathrm{cr}(\delta)}
	\vec v_e \cdot \vec v_\mathrm{PAO}\, \mathrm{d}\phi\\
	&\propto& 
	\cos(\lambda_\mathrm{PAO})\cos(\delta)\sin(\phi_\mathrm{cr})\nonumber\\
	&& \phantom{\cos(\lambda)}+ 
	\phi_\mathrm{cr} \sin(\lambda_\mathrm{PAO}) \sin(\delta) .
\ea

\begin{figure}
\center
\includegraphics[width = 0.49 \textwidth]{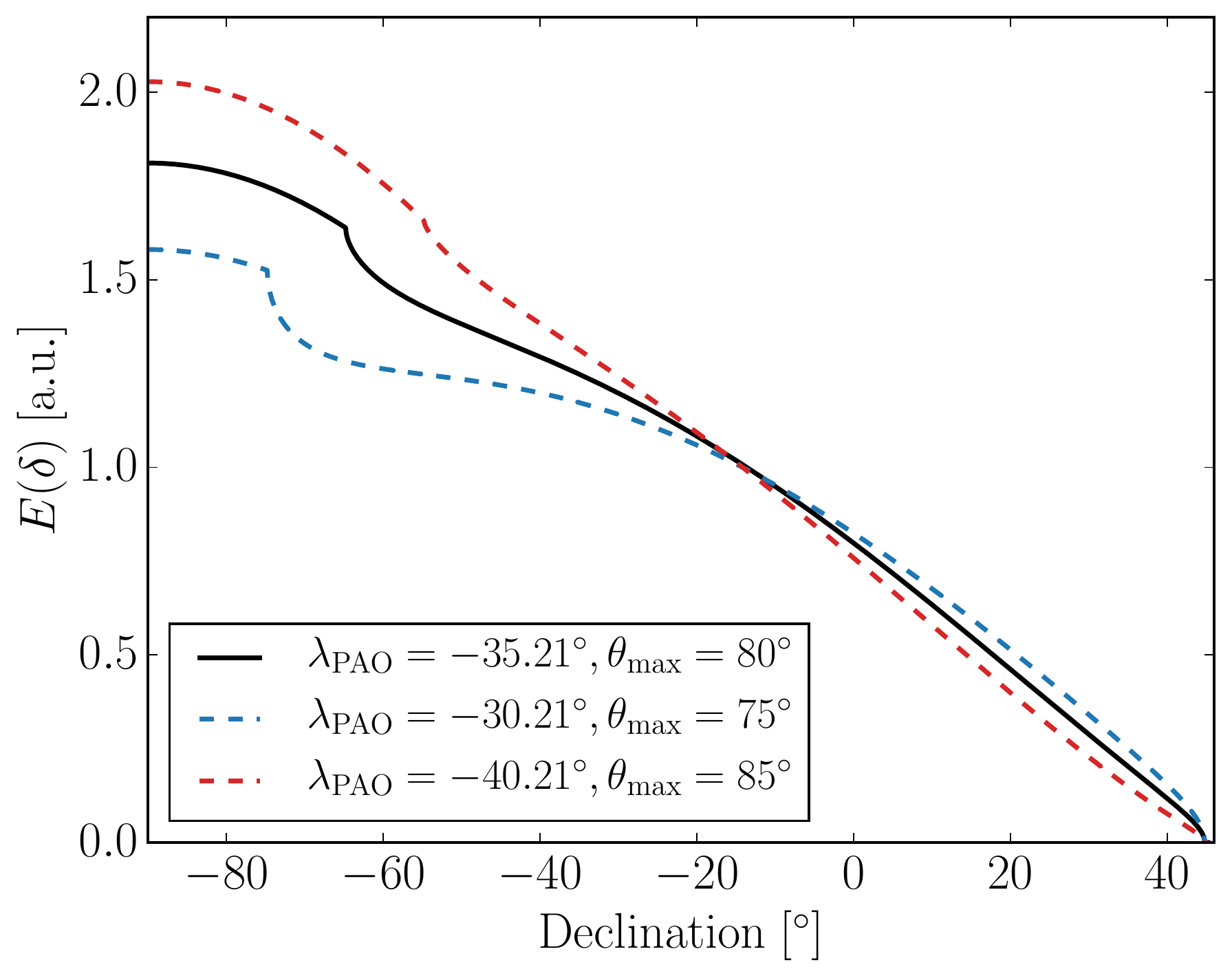}
\caption{Declination dependence of PAO exposure (black). The two dashed lines show
alternative choices of parameters we use to investigate sensitivity of our results to the
details of $E(\delta)$.}
\label{fig:exposures}
\end{figure}

For the PAO parameters, the dependence of exposure on declination is shown in
Figure~\ref{fig:exposures} (compare with Figure~S1 of \cite{Aab:2017tyv}), together with
two other choices of parameters we use to test sensitivity of our results to the details
of $E(\delta)$.

\bibliography{pa_2mrs}

\begin{thebibliography}{22}%
\makeatletter
\providecommand \@ifxundefined [1]{%
 \@ifx{#1\undefined}
}%
\providecommand \@ifnum [1]{%
 \ifnum #1\expandafter \@firstoftwo
 \else \expandafter \@secondoftwo
 \fi
}%
\providecommand \@ifx [1]{%
 \ifx #1\expandafter \@firstoftwo
 \else \expandafter \@secondoftwo
 \fi
}%
\providecommand \natexlab [1]{#1}%
\providecommand \enquote  [1]{``#1''}%
\providecommand \bibnamefont  [1]{#1}%
\providecommand \bibfnamefont [1]{#1}%
\providecommand \citenamefont [1]{#1}%
\providecommand \href@noop [0]{\@secondoftwo}%
\providecommand \href [0]{\begingroup \@sanitize@url \@href}%
\providecommand \@href[1]{\@@startlink{#1}\@@href}%
\providecommand \@@href[1]{\endgroup#1\@@endlink}%
\providecommand \@sanitize@url [0]{\catcode `\\12\catcode `\$12\catcode
  `\&12\catcode `\#12\catcode `\^12\catcode `\_12\catcode `\%12\relax}%
\providecommand \@@startlink[1]{}%
\providecommand \@@endlink[0]{}%
\providecommand \url  [0]{\begingroup\@sanitize@url \@url }%
\providecommand \@url [1]{\endgroup\@href {#1}{\urlprefix }}%
\providecommand \urlprefix  [0]{URL }%
\providecommand \Eprint [0]{\href }%
\providecommand \doibase [0]{http://dx.doi.org/}%
\providecommand \selectlanguage [0]{\@gobble}%
\providecommand \bibinfo  [0]{\@secondoftwo}%
\providecommand \bibfield  [0]{\@secondoftwo}%
\providecommand \translation [1]{[#1]}%
\providecommand \BibitemOpen [0]{}%
\providecommand \bibitemStop [0]{}%
\providecommand \bibitemNoStop [0]{.\EOS\space}%
\providecommand \EOS [0]{\spacefactor3000\relax}%
\providecommand \BibitemShut  [1]{\csname bibitem#1\endcsname}%
\let\auto@bib@innerbib\@empty
%</preamble>
\bibitem [{\citenamefont {Alves~Batista}\ \emph {et~al.}(2019)\citenamefont
  {Alves~Batista} \emph {et~al.}}]{AlvesBatista:2019tlv}%
  \BibitemOpen
  \bibfield  {author} {\bibinfo {author} {\bibfnamefont {R.}~\bibnamefont
  {Alves~Batista}} \emph {et~al.},\ }\href {\doibase 10.3389/fspas.2019.00023}
  {\bibfield  {journal} {\bibinfo  {journal} {Front. Astron. Space Sci.}\
  }\textbf {\bibinfo {volume} {6}},\ \bibinfo {pages} {23} (\bibinfo {year}
  {2019})},\ \Eprint {http://arxiv.org/abs/1903.06714} {arXiv:1903.06714
  [astro-ph.HE]} \BibitemShut {NoStop}%
\bibitem [{\citenamefont {Aab}\ \emph {et~al.}(2017)\citenamefont {Aab} \emph
  {et~al.}}]{Aab:2017tyv}%
  \BibitemOpen
  \bibfield  {author} {\bibinfo {author} {\bibfnamefont {A.}~\bibnamefont
  {Aab}} \emph {et~al.} (\bibinfo {collaboration} {Pierre Auger}),\ }\href
  {\doibase 10.1126/science.aan4338} {\bibfield  {journal} {\bibinfo  {journal}
  {Science}\ }\textbf {\bibinfo {volume} {357}},\ \bibinfo {pages} {1266}
  (\bibinfo {year} {2017})},\ \Eprint {http://arxiv.org/abs/1709.07321}
  {arXiv:1709.07321 [astro-ph.HE]} \BibitemShut {NoStop}%
\bibitem [{\citenamefont {Abbasi}\ \emph {et~al.}(2014)\citenamefont {Abbasi}
  \emph {et~al.}}]{Abbasi:2014lda}%
  \BibitemOpen
  \bibfield  {author} {\bibinfo {author} {\bibfnamefont {R.}~\bibnamefont
  {Abbasi}} \emph {et~al.} (\bibinfo {collaboration} {Telescope Array}),\
  }\href {\doibase 10.1088/2041-8205/790/2/L21} {\bibfield  {journal} {\bibinfo
   {journal} {Astrophys. J. Lett.}\ }\textbf {\bibinfo {volume} {790}},\
  \bibinfo {pages} {L21} (\bibinfo {year} {2014})},\ \Eprint
  {http://arxiv.org/abs/1404.5890} {arXiv:1404.5890 [astro-ph.HE]} \BibitemShut
  {NoStop}%
\bibitem [{\citenamefont {Aab}\ \emph {et~al.}(2018)\citenamefont {Aab} \emph
  {et~al.}}]{Aab:2018chp}%
  \BibitemOpen
  \bibfield  {author} {\bibinfo {author} {\bibfnamefont {A.}~\bibnamefont
  {Aab}} \emph {et~al.} (\bibinfo {collaboration} {Pierre Auger}),\ }\href
  {\doibase 10.3847/2041-8213/aaa66d} {\bibfield  {journal} {\bibinfo
  {journal} {Astrophys. J. Lett.}\ }\textbf {\bibinfo {volume} {853}},\
  \bibinfo {pages} {L29} (\bibinfo {year} {2018})},\ \Eprint
  {http://arxiv.org/abs/1801.06160} {arXiv:1801.06160 [astro-ph.HE]}
  \BibitemShut {NoStop}%
\bibitem [{\citenamefont {Kashti}\ and\ \citenamefont
  {Waxman}(2008)}]{Kashti:2008bw}%
  \BibitemOpen
  \bibfield  {author} {\bibinfo {author} {\bibfnamefont {T.}~\bibnamefont
  {Kashti}}\ and\ \bibinfo {author} {\bibfnamefont {E.}~\bibnamefont
  {Waxman}},\ }\href {\doibase 10.1088/1475-7516/2008/05/006} {\bibfield
  {journal} {\bibinfo  {journal} {JCAP}\ }\textbf {\bibinfo {volume} {05}},\
  \bibinfo {pages} {006} (\bibinfo {year} {2008})},\ \Eprint
  {http://arxiv.org/abs/0801.4516} {arXiv:0801.4516 [astro-ph]} \BibitemShut
  {NoStop}%
\bibitem [{\citenamefont {Koers}\ and\ \citenamefont
  {Tinyakov}(2009)}]{Koers:2008ba}%
  \BibitemOpen
  \bibfield  {author} {\bibinfo {author} {\bibfnamefont {H.~B.}\ \bibnamefont
  {Koers}}\ and\ \bibinfo {author} {\bibfnamefont {P.}~\bibnamefont
  {Tinyakov}},\ }\href {\doibase 10.1088/1475-7516/2009/04/003} {\bibfield
  {journal} {\bibinfo  {journal} {JCAP}\ }\textbf {\bibinfo {volume} {04}},\
  \bibinfo {pages} {003} (\bibinfo {year} {2009})},\ \Eprint
  {http://arxiv.org/abs/0812.0860} {arXiv:0812.0860 [astro-ph]} \BibitemShut
  {NoStop}%
\bibitem [{\citenamefont {di~Matteo}\ \emph {et~al.}(2018)\citenamefont
  {di~Matteo}, \citenamefont {Deligny}, \citenamefont {Kawata}, \citenamefont
  {de~Almeida}, \citenamefont {Mostafá}, \citenamefont {Moura~Santos},
  \citenamefont {Sagawa}, \citenamefont {Tinyakov}, \citenamefont {Tkachev},\
  and\ \citenamefont {Toshiyuki}}]{diMatteo:2018vmr}%
  \BibitemOpen
  \bibfield  {author} {\bibinfo {author} {\bibfnamefont {A.}~\bibnamefont
  {di~Matteo}}, \bibinfo {author} {\bibfnamefont {O.}~\bibnamefont {Deligny}},
  \bibinfo {author} {\bibfnamefont {K.}~\bibnamefont {Kawata}}, \bibinfo
  {author} {\bibfnamefont {R.~M.}\ \bibnamefont {de~Almeida}}, \bibinfo
  {author} {\bibfnamefont {M.}~\bibnamefont {Mostafá}}, \bibinfo {author}
  {\bibfnamefont {E.}~\bibnamefont {Moura~Santos}}, \bibinfo {author}
  {\bibfnamefont {H.}~\bibnamefont {Sagawa}}, \bibinfo {author} {\bibfnamefont
  {P.}~\bibnamefont {Tinyakov}}, \bibinfo {author} {\bibfnamefont
  {I.}~\bibnamefont {Tkachev}}, \ and\ \bibinfo {author} {\bibfnamefont
  {N.}~\bibnamefont {Toshiyuki}},\ }\href {\doibase 10.7566/JPSCP.19.011020}
  {\bibfield  {journal} {\bibinfo  {journal} {JPS Conf. Proc.}\ }\textbf
  {\bibinfo {volume} {19}},\ \bibinfo {pages} {011020} (\bibinfo {year}
  {2018})}\BibitemShut {NoStop}%
\bibitem [{\citenamefont {di~Matteo}\ \emph {et~al.}(2020)\citenamefont
  {di~Matteo} \emph {et~al.}}]{diMatteo:2020dlo}%
  \BibitemOpen
  \bibfield  {author} {\bibinfo {author} {\bibfnamefont {A.}~\bibnamefont
  {di~Matteo}} \emph {et~al.} (\bibinfo {collaboration} {Pierre Auger,
  Telescope Array}),\ }\href {\doibase 10.22323/1.358.0439} {\bibfield
  {journal} {\bibinfo  {journal} {PoS}\ }\textbf {\bibinfo {volume}
  {ICRC2019}},\ \bibinfo {pages} {439} (\bibinfo {year} {2020})},\ \Eprint
  {http://arxiv.org/abs/2001.01864} {arXiv:2001.01864 [astro-ph.HE]}
  \BibitemShut {NoStop}%
\bibitem [{\citenamefont {Abbasi}\ \emph
  {et~al.}(2018{\natexlab{a}})\citenamefont {Abbasi} \emph
  {et~al.}}]{Abbasi:2018tqo}%
  \BibitemOpen
  \bibfield  {author} {\bibinfo {author} {\bibfnamefont {R.}~\bibnamefont
  {Abbasi}} \emph {et~al.} (\bibinfo {collaboration} {Telescope Array}),\
  }\href {\doibase 10.3847/2041-8213/aaebf9} {\bibfield  {journal} {\bibinfo
  {journal} {Astrophys. J. Lett.}\ }\textbf {\bibinfo {volume} {867}},\
  \bibinfo {pages} {L27} (\bibinfo {year} {2018}{\natexlab{a}})},\ \Eprint
  {http://arxiv.org/abs/1809.01573} {arXiv:1809.01573 [astro-ph.HE]}
  \BibitemShut {NoStop}%
\bibitem [{\citenamefont {Tinyakov}(2018)}]{Tinyakov:2018hfg}%
  \BibitemOpen
  \bibfield  {author} {\bibinfo {author} {\bibfnamefont {P.}~\bibnamefont
  {Tinyakov}} (\bibinfo {collaboration} {Telescope Array}),\ }\href {\doibase
  10.7566/JPSCP.19.011019} {\bibfield  {journal} {\bibinfo  {journal} {JPS
  Conf. Proc.}\ }\textbf {\bibinfo {volume} {19}},\ \bibinfo {pages} {011019}
  (\bibinfo {year} {2018})}\BibitemShut {NoStop}%
\bibitem [{\citenamefont {Oikonomou}\ \emph {et~al.}(2013)\citenamefont
  {Oikonomou}, \citenamefont {Connolly}, \citenamefont {Abdalla}, \citenamefont
  {Lahav}, \citenamefont {Thomas}, \citenamefont {Waters},\ and\ \citenamefont
  {Waxman}}]{Oikonomou:2012ef}%
  \BibitemOpen
  \bibfield  {author} {\bibinfo {author} {\bibfnamefont {F.}~\bibnamefont
  {Oikonomou}}, \bibinfo {author} {\bibfnamefont {A.}~\bibnamefont {Connolly}},
  \bibinfo {author} {\bibfnamefont {F.~B.}\ \bibnamefont {Abdalla}}, \bibinfo
  {author} {\bibfnamefont {O.}~\bibnamefont {Lahav}}, \bibinfo {author}
  {\bibfnamefont {S.~A.}\ \bibnamefont {Thomas}}, \bibinfo {author}
  {\bibfnamefont {D.}~\bibnamefont {Waters}}, \ and\ \bibinfo {author}
  {\bibfnamefont {E.}~\bibnamefont {Waxman}},\ }\href {\doibase
  10.1088/1475-7516/2013/05/015} {\bibfield  {journal} {\bibinfo  {journal}
  {JCAP}\ }\textbf {\bibinfo {volume} {05}},\ \bibinfo {pages} {015} (\bibinfo
  {year} {2013})},\ \Eprint {http://arxiv.org/abs/1207.4043} {arXiv:1207.4043
  [astro-ph.HE]} \BibitemShut {NoStop}%
\bibitem [{\citenamefont {Aab}\ \emph {et~al.}(2015{\natexlab{a}})\citenamefont
  {Aab} \emph {et~al.}}]{PierreAuger:2014yba}%
  \BibitemOpen
  \bibfield  {author} {\bibinfo {author} {\bibfnamefont {A.}~\bibnamefont
  {Aab}} \emph {et~al.} (\bibinfo {collaboration} {Pierre Auger}),\ }\href
  {\doibase 10.1088/0004-637X/804/1/15} {\bibfield  {journal} {\bibinfo
  {journal} {Astrophys. J.}\ }\textbf {\bibinfo {volume} {804}},\ \bibinfo
  {pages} {15} (\bibinfo {year} {2015}{\natexlab{a}})},\ \Eprint
  {http://arxiv.org/abs/1411.6111} {arXiv:1411.6111 [astro-ph.HE]} \BibitemShut
  {NoStop}%
\bibitem [{\citenamefont {Abbasi}\ \emph
  {et~al.}(2018{\natexlab{b}})\citenamefont {Abbasi} \emph
  {et~al.}}]{Abbasi:2018qlh}%
  \BibitemOpen
  \bibfield  {author} {\bibinfo {author} {\bibfnamefont {R.}~\bibnamefont
  {Abbasi}} \emph {et~al.} (\bibinfo {collaboration} {Telescope Array}),\
  }\href {\doibase 10.3847/1538-4357/aac9c8} {\bibfield  {journal} {\bibinfo
  {journal} {Astrophys. J.}\ }\textbf {\bibinfo {volume} {862}},\ \bibinfo
  {pages} {91} (\bibinfo {year} {2018}{\natexlab{b}})},\ \Eprint
  {http://arxiv.org/abs/1802.05003} {arXiv:1802.05003 [astro-ph.HE]}
  \BibitemShut {NoStop}%
\bibitem [{\citenamefont {Denton}\ and\ \citenamefont
  {Weiler}(2015)}]{Denton:2014hfa}%
  \BibitemOpen
  \bibfield  {author} {\bibinfo {author} {\bibfnamefont {P.~B.}\ \bibnamefont
  {Denton}}\ and\ \bibinfo {author} {\bibfnamefont {T.~J.}\ \bibnamefont
  {Weiler}},\ }\href {\doibase 10.1088/0004-637X/802/1/25} {\bibfield
  {journal} {\bibinfo  {journal} {Astrophys. J.}\ }\textbf {\bibinfo {volume}
  {802}},\ \bibinfo {pages} {25} (\bibinfo {year} {2015})},\ \Eprint
  {http://arxiv.org/abs/1409.0883} {arXiv:1409.0883 [astro-ph.HE]} \BibitemShut
  {NoStop}%
\bibitem [{\citenamefont {Ahlers}\ \emph {et~al.}(2018)\citenamefont {Ahlers},
  \citenamefont {Denton},\ and\ \citenamefont {Rameez}}]{Ahlers:2017wpb}%
  \BibitemOpen
  \bibfield  {author} {\bibinfo {author} {\bibfnamefont {M.}~\bibnamefont
  {Ahlers}}, \bibinfo {author} {\bibfnamefont {P.}~\bibnamefont {Denton}}, \
  and\ \bibinfo {author} {\bibfnamefont {M.}~\bibnamefont {Rameez}},\ }\href
  {\doibase 10.22323/1.301.0282} {\bibfield  {journal} {\bibinfo  {journal}
  {PoS}\ }\textbf {\bibinfo {volume} {ICRC2017}},\ \bibinfo {pages} {282}
  (\bibinfo {year} {2018})}\BibitemShut {NoStop}%
\bibitem [{\citenamefont {Urban}\ \emph {et~al.}(2020)\citenamefont {Urban},
  \citenamefont {Camera},\ and\ \citenamefont {Alonso}}]{Urban:2020szk}%
  \BibitemOpen
  \bibfield  {author} {\bibinfo {author} {\bibfnamefont {F.~R.}\ \bibnamefont
  {Urban}}, \bibinfo {author} {\bibfnamefont {S.}~\bibnamefont {Camera}}, \
  and\ \bibinfo {author} {\bibfnamefont {D.}~\bibnamefont {Alonso}},\
  }\href@noop {} {\  (\bibinfo {year} {2020})},\ \Eprint
  {http://arxiv.org/abs/2005.00244} {arXiv:2005.00244 [astro-ph.HE]}
  \BibitemShut {NoStop}%
\bibitem [{\citenamefont {Aab}\ \emph {et~al.}(2015{\natexlab{b}})\citenamefont
  {Aab} \emph {et~al.}}]{ThePierreAuger:2015rma}%
  \BibitemOpen
  \bibfield  {author} {\bibinfo {author} {\bibfnamefont {A.}~\bibnamefont
  {Aab}} \emph {et~al.} (\bibinfo {collaboration} {Pierre Auger}),\ }\href
  {\doibase 10.1016/j.nima.2015.06.058} {\bibfield  {journal} {\bibinfo
  {journal} {Nucl. Instrum. Meth. A}\ }\textbf {\bibinfo {volume} {798}},\
  \bibinfo {pages} {172} (\bibinfo {year} {2015}{\natexlab{b}})},\ \Eprint
  {http://arxiv.org/abs/1502.01323} {arXiv:1502.01323 [astro-ph.IM]}
  \BibitemShut {NoStop}%
\bibitem [{\citenamefont {Huchra}\ \emph {et~al.}(2012)\citenamefont {Huchra}
  \emph {et~al.}}]{Huchra:2011ii}%
  \BibitemOpen
  \bibfield  {author} {\bibinfo {author} {\bibfnamefont {J.~P.}\ \bibnamefont
  {Huchra}} \emph {et~al.},\ }\href {\doibase 10.1088/0067-0049/199/2/26}
  {\bibfield  {journal} {\bibinfo  {journal} {Astrophys. J. Suppl.}\ }\textbf
  {\bibinfo {volume} {199}},\ \bibinfo {pages} {26} (\bibinfo {year} {2012})},\
  \Eprint {http://arxiv.org/abs/1108.0669} {arXiv:1108.0669 [astro-ph.CO]}
  \BibitemShut {NoStop}%
\bibitem [{\citenamefont {Gorski}\ \emph {et~al.}(2005)\citenamefont {Gorski},
  \citenamefont {Hivon}, \citenamefont {Banday}, \citenamefont {Wandelt},
  \citenamefont {Hansen}, \citenamefont {Reinecke},\ and\ \citenamefont
  {Bartelman}}]{Gorski:2004by}%
  \BibitemOpen
  \bibfield  {author} {\bibinfo {author} {\bibfnamefont {K.}~\bibnamefont
  {Gorski}}, \bibinfo {author} {\bibfnamefont {E.}~\bibnamefont {Hivon}},
  \bibinfo {author} {\bibfnamefont {A.}~\bibnamefont {Banday}}, \bibinfo
  {author} {\bibfnamefont {B.}~\bibnamefont {Wandelt}}, \bibinfo {author}
  {\bibfnamefont {F.}~\bibnamefont {Hansen}}, \bibinfo {author} {\bibfnamefont
  {M.}~\bibnamefont {Reinecke}}, \ and\ \bibinfo {author} {\bibfnamefont
  {M.}~\bibnamefont {Bartelman}},\ }\href {\doibase 10.1086/427976} {\bibfield
  {journal} {\bibinfo  {journal} {Astrophys. J.}\ }\textbf {\bibinfo {volume}
  {622}},\ \bibinfo {pages} {759} (\bibinfo {year} {2005})},\ \Eprint
  {http://arxiv.org/abs/astro-ph/0409513} {arXiv:astro-ph/0409513} \BibitemShut
  {NoStop}%
\bibitem [{\citenamefont {Hivon}\ \emph {et~al.}(2002)\citenamefont {Hivon},
  \citenamefont {Gorski}, \citenamefont {Netterfield}, \citenamefont {Crill},
  \citenamefont {Prunet},\ and\ \citenamefont {Hansen}}]{Hivon:2001jp}%
  \BibitemOpen
  \bibfield  {author} {\bibinfo {author} {\bibfnamefont {E.}~\bibnamefont
  {Hivon}}, \bibinfo {author} {\bibfnamefont {K.}~\bibnamefont {Gorski}},
  \bibinfo {author} {\bibfnamefont {C.}~\bibnamefont {Netterfield}}, \bibinfo
  {author} {\bibfnamefont {B.}~\bibnamefont {Crill}}, \bibinfo {author}
  {\bibfnamefont {S.}~\bibnamefont {Prunet}}, \ and\ \bibinfo {author}
  {\bibfnamefont {F.}~\bibnamefont {Hansen}},\ }\href {\doibase 10.1086/338126}
  {\bibfield  {journal} {\bibinfo  {journal} {Astrophys. J.}\ }\textbf
  {\bibinfo {volume} {567}},\ \bibinfo {pages} {2} (\bibinfo {year} {2002})},\
  \Eprint {http://arxiv.org/abs/astro-ph/0105302} {arXiv:astro-ph/0105302}
  \BibitemShut {NoStop}%
\bibitem [{\citenamefont {Chon}\ \emph {et~al.}(2004)\citenamefont {Chon},
  \citenamefont {Challinor}, \citenamefont {Prunet}, \citenamefont {Hivon},\
  and\ \citenamefont {Szapudi}}]{Chon:2003gx}%
  \BibitemOpen
  \bibfield  {author} {\bibinfo {author} {\bibfnamefont {G.}~\bibnamefont
  {Chon}}, \bibinfo {author} {\bibfnamefont {A.}~\bibnamefont {Challinor}},
  \bibinfo {author} {\bibfnamefont {S.}~\bibnamefont {Prunet}}, \bibinfo
  {author} {\bibfnamefont {E.}~\bibnamefont {Hivon}}, \ and\ \bibinfo {author}
  {\bibfnamefont {I.}~\bibnamefont {Szapudi}},\ }\href {\doibase
  10.1111/j.1365-2966.2004.07737.x} {\bibfield  {journal} {\bibinfo  {journal}
  {Mon. Not. Roy. Astron. Soc.}\ }\textbf {\bibinfo {volume} {350}},\ \bibinfo
  {pages} {914} (\bibinfo {year} {2004})},\ \Eprint
  {http://arxiv.org/abs/astro-ph/0303414} {arXiv:astro-ph/0303414} \BibitemShut
  {NoStop}%
\bibitem [{\citenamefont {Ando}\ \emph {et~al.}(2018)\citenamefont {Ando},
  \citenamefont {Benoit-Lévy},\ and\ \citenamefont {Komatsu}}]{Ando:2017wff}%
  \BibitemOpen
  \bibfield  {author} {\bibinfo {author} {\bibfnamefont {S.}~\bibnamefont
  {Ando}}, \bibinfo {author} {\bibfnamefont {A.}~\bibnamefont {Benoit-Lévy}},
  \ and\ \bibinfo {author} {\bibfnamefont {E.}~\bibnamefont {Komatsu}},\ }\href
  {\doibase 10.1093/mnras/stx2634} {\bibfield  {journal} {\bibinfo  {journal}
  {Mon. Not. Roy. Astron. Soc.}\ }\textbf {\bibinfo {volume} {473}},\ \bibinfo
  {pages} {4318} (\bibinfo {year} {2018})},\ \Eprint
  {http://arxiv.org/abs/1706.05422} {arXiv:1706.05422 [astro-ph.CO]}
  \BibitemShut {NoStop}%
\end{thebibliography}%

\end{document}